\documentclass[12pt,preprint]{aastex}

\slugcomment{Accepted by ApJ Letters} 
 
\shorttitle{RGB-bump in Sgr dSph} 
\shortauthors{Monaco et al.} 
 
\begin{document}

\title{First detection of the RGB-bump in the Sagittarius
dSph\altaffilmark{1,2}} 

\altaffiltext{1}{Based on observations made with the European Southern 
Observatory telescopes, using the Wide Field Imager, as part of the 
observing program 65.L-0463. Also based on data obtained from the 
ESO/ST-ECF Science Archive Facility.} 
\altaffiltext{2}{This publication makes use of data products from the 
Two Micron All Sky Survey, which is a joint project of the University of 
Massachusetts and the Infrared Processing and Analysis Center/California 
Institute of Technology, funded by the National Aeronautics and Space 
Administration and the National Science Foundation} 
 
\author{L. Monaco} 
\affil{Dipartimento di Astronomia, Universit\`a di Bologna, 
 via Ranzani,1 40127 Bologna, ITALY} 
\email{monaco@bo.astro.it} 
 
\author{F. R. Ferraro, M. Bellazzini} 
\affil{INAF - Osservatorio Astronomico di Bologna, 
 via Ranzani,1 40127 Bologna, ITALY} 
\email{ferraro,bellazzini@bo.astro.it} 
 
\and 
 
\author{E. Pancino} 
\affil{Dipartimento di Astronomia, Universit\`a di Bologna, 
 via Ranzani,1 40127 Bologna, ITALY} 
\email{pancino@bo.astro.it}

 
\begin{abstract} 
 
We present V, I photometry of the Sagittarius Dwarf Spheroidal galaxy 
(Sgr) for a region of $\sim 1^{\circ} \times 1^{\circ}$, centered on 
the globular cluster M~54. This catalog is the largest database of 
stars ($\sim$500,000) ever obtained for this galaxy. The wide area 
covered allows us to measure for the first time the position of the 
RGB-bump, a feature that has been identified in most Galactic globular 
clusters and only recently in a few galaxies of the Local Group. The 
presence of a single-peaked bump in the RGB differential Luminosity 
Function confirms that there is a dominant population in Sgr (Pop~A). 

The photometric properties of the Pop~A RGB and the position of the RGB
bump have been used to constrain the range of possible ages and 
metallicities of this population. The most likely solution lies in the
range $-0.6< [M/H]\le -0.4$ and $4~Gyr\le age\le 8~Gyr$. 

\end{abstract} 
 

 
\section{Introduction} 
 
The Sagittarius dwarf spheroidal (Sgr dSph) galaxy \citep{igi,iba97} is
an {\em in vivo} example of an accretion/disruption event of a Galactic
satellite \citep{new02}: a process that may have been one of the main
mechanisms for the assembly of the Galactic halo \citep{sz78,cote}.
Hence, the study of the stellar content, star formation history and
chemical evolution \cite[see][for references]{ls00,b99b,sm02} of this
disrupting system is a fundamental step in order to reconstruct the
evolutionary history of the Milky Way.

As part of a long-term project devoted to understand the origin and
the evolution of the {\em building blocks} of the Galactic Halo, we 
present here the first result of a new wide-field V,I photometry of
the Sgr in a region ($\sim$1$^{\circ}\times $1$^{\circ}$) around the 
globular cluster M~54. Almost 500,000 stars have been measured in the 
area, allowing us to unambiguously identify, for the first time, the 
RGB-bump in the Sgr. This feature was identified in a relevant number 
of Galactic globular clusters \citep{fp90,f99,z99} and only in a few 
galaxies in the Local Group: Sculptor \citep{maj99}, Sextans 
\citep{b01} and Ursa Minor \citep{b02}. 

The RGB-bump magnitude is mainly driven by the metal content and, to a
lesser extent, by the age of the population. In this {\em Letter} we
use this feature to constrain the metallicty and age of the main 
stellar population in the Sgr galaxy (Pop~A). Such constraints are 
particularly useful in the case of the Sgr dSph, where the strong 
foreground contamination makes a clear-cut interpretation of the Color 
Magnitude Diagram (CMD) more difficult \citep[see][hereafter B99 and 
LS00, respectively]{b99b,ls00} and where some inconsistency between
the photometric and spectroscopic metallicity estimates has emerged 
\citep{co01,bo00,cse01,sm02}.

 
\section{Observations and results} 
 
The data presented here consist of a sample of images obtained at the 
2.2m ESO/MPI telescope at la Silla, Chile, using the {\it Wide Field 
Imager} (WFI), a mosaic of eight $2048\times4096$ pixels CCD arrays. 
The scale is $0.238"/px$, giving a total field of view of $\sim 34' 
\times 33'$. A set of V and I images in the vicinity of the Globular 
Cluster M~54 were secured during an observing run on 7-8 July 2000, 
with typical exposure times ranging from 10 to 400 sec. To complement 
this data-set, we retrieved a set of public WFI images from the 
ESO/ST-ECF Science Archive Facility. The public data were originally 
obtained in the framework of the ESO pre-FLAMES survey, covering a wide 
region of $\sim 1^{\circ} \times 1^{\circ}$ around M~54. 
 
The entire data-set was homogeneously treated: in short, the raw
images were corrected for bias and flat-field using standard 
IRAF\footnote{IRAF is distributed by the National Optical Astronomy 
Observatories, which is operated by the association of Universities
for Research in Astronomy, Inc., under contract with the National
Science Foundation.} procedures, within the {\it noao.mscred} package.
The relative photometry was performed with the Point Spread Function
(PSF) fitting package DoPhot \citep{dophot}. The stars were searched
independently on each image using a spatially variable PSF. We used the
large overlapping regions between adjacent WFI pointings to transform
the instrumental magnitudes into a homogeneous system, eventually
correcting for the zero-point differences between chips and obtaining
a final catalog of homogeneous magnitudes and coordinates. Finally,
the photometric calibration was performed using more than 1000 stars
in common with the \citet{sl95} catalog (hereafter SL95). 

\begin{figure} 
\plotone{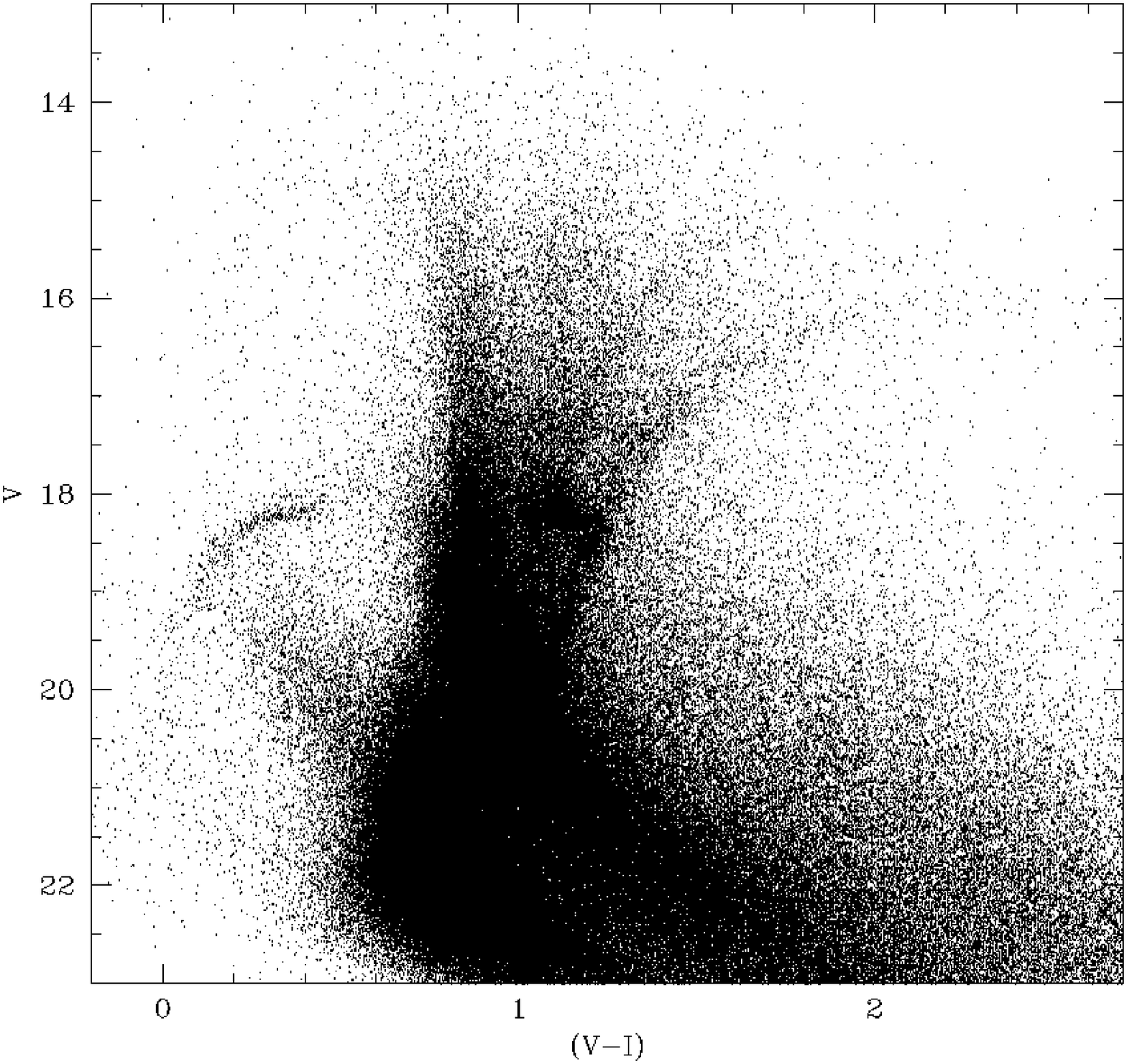}
\caption{Color-magnitude diagram of 493,000 stars in the Sgr dSph,
covering a region of about $1^{\circ}\times$1$^{\circ}$ around the
globular cluster M~54. Three main populations, i.e., the field stars,
M~54 and the Sgr, are clearly visible.} 
\label{f1} 
\end{figure}

The $(V,V-I)$ CMD for $\sim$493,000 stars measured in the global field 
of view is shown in Figure~\ref{f1}. At least three main populations 
can be distinguished: 

({\it i}) the RGB of the Sgr's metal-rich population at the red side of 
the diagram, extending to a very red color, $(V-I)\sim 2.7$, with the 
corresponding, well populated red HB clump at $V\sim18.22$. The RGB 
bump of the Sgr can easily be seen as a clump of stars along the RGB at 
$V\sim 18.5$ and $(V-I)\sim 1.2$. 
 
({\it ii}) a second, much steeper RGB, barely visible at $(V,V-I)\sim 
(16.5,1.3)$, belonging to the much metal-poorer population of M~54. Its blue 
HB is clearly seen at $(V,V-I)=(18.2,0.2)$. 
 
({\it iii}) the field population, a nearly vertical sequence around 
$V-I=0.8$ and, at slighly redder colors (V-I$\sim$1.1), the evolved Milky Way
bulge/disk stars. 

The photometric properties of the Sub Giant Branch and Main Sequence 
Turn Off (TO) region (at $V\sim 21.5$) will be discussed in a 
forthcoming paper.

 
\section{The RGB-bump} 
 
The so-called {\it RGB-bump} is an evolutionary feature predicted by
the theoretical models of stellar evolution \citep{thomas,iben}. The 
physical origin of the RGB-bump is well known: in the stellar interior 
during the first ascent of the giant branch, the H-burning shell is 
moving constantly outwards until it reaches the discontinuity in the H 
radial profile, left by the innermost penetration of the convective 
envelope inside the star. At the discontinuity, the luminosity of the 
star drops for a while, until the shell adapts to the new environment, 
and then it rises again in a regime of constant H content. 
Observationally, the RGB-bump appears as a peak in the differential 
luminosity function (LF), due to the non-monotonic evolution of the 
star in luminosity. It appears also as a jump and a change in slope of 
the integrated LF, due to the change in the evolutionary rate, caused 
by the sudden change of molecular weight of the material in which the 
H-burning shell is moving. According to \citet{fp90}, the change in
the slope of the integrated luminosity function is the safest way to 
identify the location of the bump, since it makes use of stars 
contained in several magnitude bins \cite[see also][F99 
hereafter]{f99}. 

Since its first detection in 47~Tuc \citep{king}, the RGB-bump has
been identified in many globular clusters \citep[see][]{fp90,f99,z99},
and only recently in a few Local Group dwarf galaxies
\citep{maj99,b01,b02}. In complex stellar populations, such as
galaxies, the RGB-bump appears sometimes as a double-peaked feature in
the differential LF (see the case of Sculptor and Sextans). Since the
RGB-bump magnitude depends on the metal content and age of the
population (F99), the double-peak has been interpreted as a signature
of the existence of two main stellar populations in the galaxy stellar
mix \citep{maj99,b01}. 

\begin{figure} 
\plotone{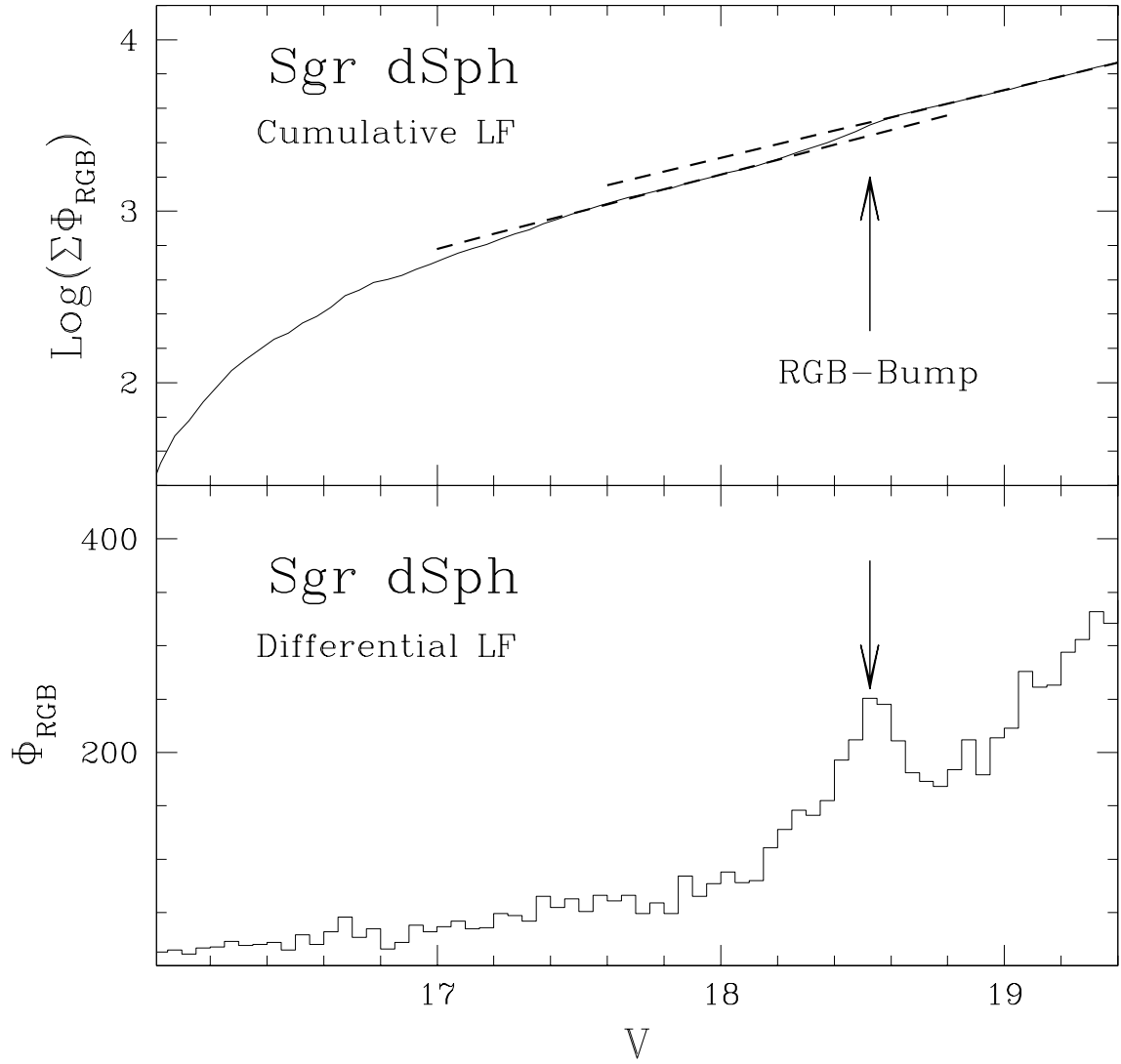} 
\caption{Differential (lower panel) and cumulative (upper panel) luminosity 
functions for the stars selected in the region of the RGB of the Sgr. 
The arrows indicate the location of the RGB-bump at V=18.55$\pm$0.05.} 
\label{f2} 
\end{figure}

In Figure~\ref{f2}, we show the differential (lower panel) and 
cumulative (upper panel) luminosity functions for the RGB of Sgr, 
selected from Figure~\ref{f1}. As can be seen, more than 5000 stars 
have been counted along the brighter ($V<19$) portion of the Sgr RGB. 
The RGB-bump is clearly identified in both panels of figure~\ref{f2}: 
it is located at V$^{bump}$=18.55$\pm$0.05. {\em This is the first 
detection of such a feature in the Sgr dSph}. In the simplest scenario 
the presence of a single-peaked bump in the differential LF suggest the 
existence of a dominant stellar population, relatively homogeneous 
in metallicity and age, in the inner regions of the Sgr galaxy\footnote{It 
is important to recall that the present analysis refers to the center of the 
Sgr dSph, since a significant population gradient is present in this 
galaxy. The outer region hosts a more metal poor dominant population 
\cite[see B99, LS00,][and references therein]{ala01}.} 
\citep{marc,b99b,cbc,dolphin}.

 
\section{Boxing the metallicity and age of the Sgr main population} 
 
Several authors \citep{marc,b99b,ls00} have found evidence that the 
dominant population \citep[Pop~A, in the nomenclature introduced 
by][]{b99b} in the central part of Sgr is similar to the globular
cluster Terzan~7 (a member of the Sgr galaxy), i.e., significantly
younger than the typical Galactic globular but still several Gyr old
\citep{buo}. 

\begin{figure} 
\plotone{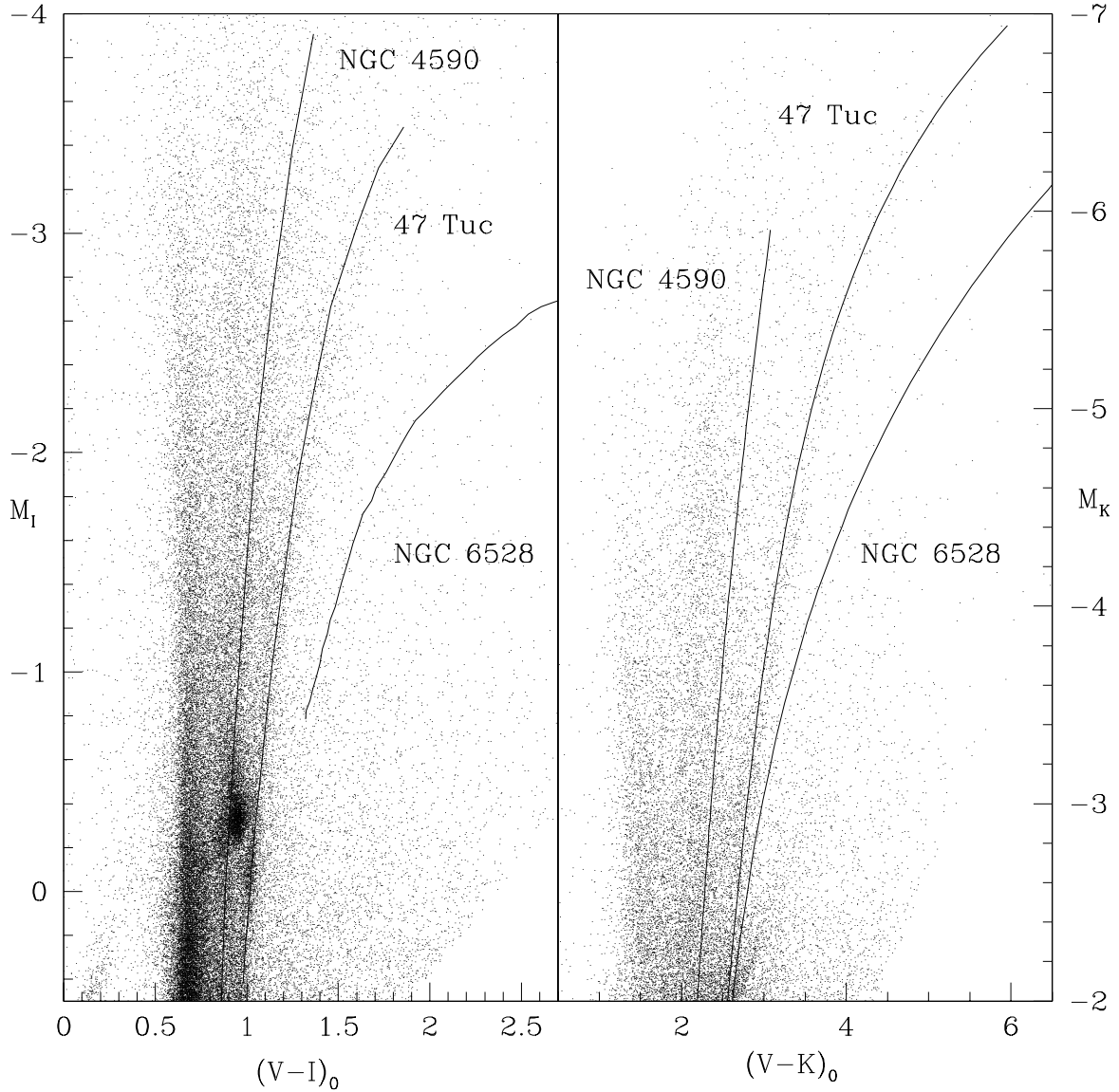} 
\caption{Color-magnitude diagrams in the absolute M$_{I}$ vs
(V-I)$_{0}$ (left panel) and M$_{K}$ vs (V-K)$_{0}$ (right panel)
planes. We adopted the distance and reddening estimates by LS00, and
the reddening laws by \citet{dean} for the V and I passbands, and by
\citet{sama} for the infrared colors. The ridge-lines of three template
globular clusters of different metallicities are overplotted (see text
for references). The assumed reddening and distance moduli for these
clusters are from F99.} 
\label{f3}  
\end{figure}

A first hint on the metal content of Pop~A can be obtained by comparing
the position and the shape of the Sgr RGB with the ridge lines of a set
of template globular clusters. Such a comparison, in the $(I,V-I)$ and
in the $(K,V-K)$ planes, is shown in Figure~\ref{f3}. The $(K,V-K)$ CMD
has been obtained by combining the catalog presented here with the Two
Micron All Sky Survey (2MASS)\footnote{See:
http://www.ipac.caltech.edu/2mass} J,H,K$_S$ dataset. We used the color
transformations computed by \citet{car01} to convert the 2MASS
magnitudes into the standard system \citep{elias}. The comparison with
the mean ridge lines of three reference globulars of different
metallicity \citep[NGC4590, 47Tuc, and NGC6528
from][respectively]{orto,sav,wa} is shown in Figure~\ref{f3}. In both
planes the ridge line of 47~Tuc ([M/H]$\simeq -0.6$) reproduces well
the RGB of Pop~A. 

Now, we can use the RGB-bump to obtain an additional, independent 
estimate of the mean metallicity of Pop~A. In particular, we use the 
reddening-free and distance-independent parameter 
$\Delta$V$^{bump}_{HB}$ (see F99), namely the difference between the V 
magnitude of the RGB-bump and the Zero Age Horizontal Branch (ZAHB) 
level. To evaluate $V_{ZAHB}$ we followed the procedure described in 
details in F99: we found V$_{ZAHB}$=18.33$\pm$0.07 and finally $\Delta 
V^{bump}_{HB}=18.55-18.33=0.22\pm 0.10 $. Using the F99 database, we 
obtain a relation for the global metallicity as a function of $\Delta 
V^{bump}_{HB}$: $$[M/H] = -0.522(\Delta V^{bump}_{HB})^2+1.115\Delta
V^{bump}_{HB}-0.862 $$ $$~~r.m.s=0.07$$ From this equation we obtain
$[M/H]=-0.64 \pm 0.12$ for Pop~A in the Sgr dSph. 

The similarities between the Sgr Pop~A and 47~Tuc allow us to perform a 
more detailed differential comparison between these two populations. In 
doing this, we shifted the CMD of 47~Tuc \cite[from the dataset 
of][]{ros00} by 4.22 magnitudes to match the Pop~A red HB. As a result, 
we find that the RGB-bump of Pop~A is significantly brighter than that 
of 47~Tuc by $\Delta V^{bump}_{\rm{47~Tuc-Sgr}} = 0.12 \pm 0.07$~mag. 
The result is confirmed by the comparison of the $\Delta V^{bump}_{HB}$ 
above with the $\Delta V^{bump}_{HB} = 0.33 \pm 0.1$ listed by F99 for 
47~Tuc, giving $\Delta V^{bump}_{\rm{47~Tuc-Sgr}} = 0.11 \pm 0.15$ mag. 
If the metallicity of the two populations is the same, then such a difference 
in the RGB-bump position corresponds to an 
age difference of $\simeq 5$ Gyr (since $\frac{\Delta M_V^{bump}} 
{\Delta t(Gyr)} =0.024$, \citet{cs97}), Pop~A being  younger than 47~Tuc. 
This finding is in excellent agreement with the results obtained by B99 
and LS00 from the study of the Main Sequence Turn Off (TO) of the Sgr 
galaxy. 
 
On the other hand, since at a fixed metallicity a younger age implies 
bluer RGB colors, we have to conclude that the mean metallicity of 
Pop~A should be higher than that of 47~Tuc \cite[see, e.g.][]{co01}. 
This is in good agreement with the most recent spectroscopic measures: 
eight of the fourteen Sgr stars observed by \citet{sm02} have $-0.6\le 
[Fe/H] \le -0.3$ and $<[\alpha/Fe]> \simeq 0.0$ (hence $[Fe/H]=[M/H]$).
Since it is reasonable to assume that these stars are representative of
Pop~A, we can deduce that the mean metallicity for this population is
$[M/H] \le -0.4$. 

\begin{figure} 
\plotone{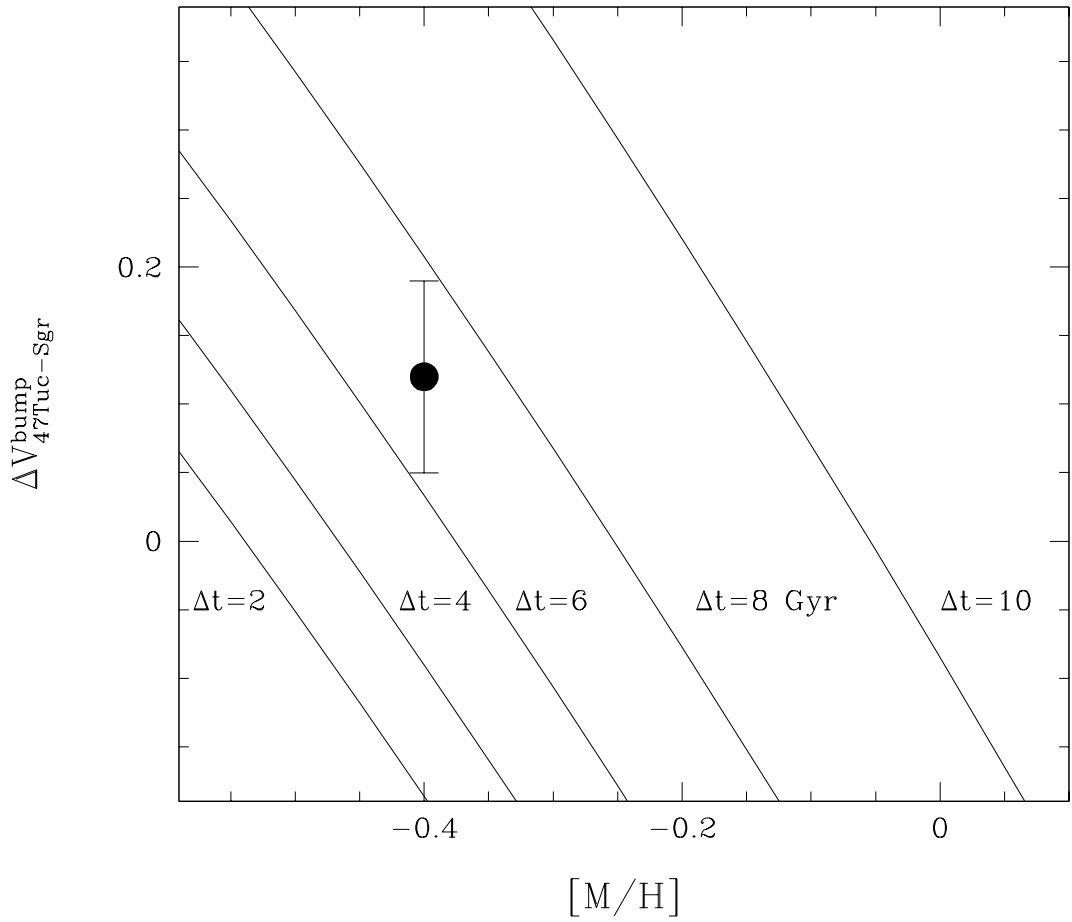}
 
\caption{The observed difference in magnitude between the RGB-bump in
the Sgr Pop~A with respect to 47~Tuc is plotted as a large filled dot.
The solid lines are theoretical isochrones computed using equation~3 by
F99, for various age differences. A global metallicity of $[M/H]=-0.6$
and an age of 12~Gyr was assumed for 47~Tuc.} 

\label{f4} 
\end{figure}

As can be seen from Figure~\ref{f4}, by adopting this metallicity the 
observed $\Delta V^{bump}_{\rm{47~Tuc-Sgr}}$ suggests an age difference
of $\sim 7$ Gyr. In a scale in which the age of 47~Tuc is 12 Gyr
\citep{carr00}, the absolute age of Pop~A turns out to be $\sim 5$ Gyr,
in good agreement with the results by LS00.

 
\section{Discussion} 
 
The growing wealth of independent observational material about the Sgr 
dSph is beginning to provide a self-consistent view of the properties 
of this galaxy. For instance, the presence of a metal poor population 
($-2.0\la [Fe/H] \la -1.4$), first suggested by B99, has been confirmed
by further photometric studies \cite[LS00,][]{co01}, by the analysis of
the pulsational properties of the RR Lyrae variables \citep{cse01} and
also by the spectroscopic survey by \citet{sm02}. Moreover, the
existence of a significant metallicity (and/or age) spatial gradient,
with the stars in the outer regions being more metal deficient than
those near the center of the galaxy, has been put into evidence by
different authors \cite[B99, LS00][]{ala01}. 

Conversely, some reason of concern was provided by the estimate of the 
mean metallicity of the stellar population that dominates the inner 
region of the galaxy. While photometric estimates were tipically in the
range $-1.0\la [Fe/H]\la -0.5$, the first high-resolution spectra of
two member stars by \citet{bo00} suggested a much higher metal content
($[Fe/H]\simeq -0.2$). The larger sample provided by the spectroscopic
analysis of \citet{sm02} suggests that, while stars having
$[Fe/H]\simeq -0.2$ are indeed present in Sgr, they do not represent
the dominant population. The majority of stars in the \citet{sm02}
sample has $[\alpha/Fe]\simeq 0.0$ (and thus [Fe/H]=[M/H]) and $-0.6\le
[Fe/H] \le -0.3$. Furthermore, previous photometric estimates based on
the comparison of the RGB with template globular clusters, neglected
the effects of age and the difference in the $\alpha$-elements
enhancement, as correctly pointed out by \citet{co01}. 

In this {\em Letter} we have revisited the problem of the metallicity 
and age of the Sgr main population, taking into account all of the 
above considerations and adding to the usual photometric metallicity 
indices (i.e., the position and the shape of the RGB) an additional 
constraint provided by the position of the RGB-bump. The main results 
can be summarized as follows: 

\begin{enumerate} 
 
\item The mere detection of a single-peaked bump in the differential LF 
of the RGB confirms the existence of a dominant population (Pop~A) in 
the Sgr dSph stellar mix. 
 
\item The Pop~A RGB is well fitted by the RGB ridge-line of the 
globular cluster 47~Tuc, that has $[M/H]\simeq -0.6$. However, a 
significant difference in the RGB-bump luminosity between Sgr and 
47~Tuc has been measured. This fact suggest that Pop A is several Gyr 
younger than 47~Tuc, in good agreement with previous results based on 
the comparison of the Main Sequence Turn Off's (B99, LS00). Hence, 
$[M/H]\simeq -0.6$ has to be considered as a {\em lower limit} to the 
mean metallicity of Pop~A. 
 
\item A full self-consistency among the spectroscopic and photometric 
constraints (including the RGB-bump) is achieved if a mean metallicity 
of $-0.6< [M/H]\le -0.4$ and a mean age of $7\ge age\ge 4$ Gyr are 
assumed for Pop~A, in good agreement with the results by LS00. Ages 
younger than $\sim 4$ Gyr are {\em excluded} by the observed morphology 
of the TO (see Figure~11 by B99 and Figure~16 by LS00). The detailed 
discussion of this region of the CMD will be the subject of a 
forthcoming paper in preparation. 
 
\end{enumerate}

 
\acknowledgments 
 
This research is partially supported by MIUR (Ministero della 
Istruzione, dell'Universit\`a e della Ricerca) and ASI (Agenzia 
spaziale Italiana). Part of the data analysis has been performed using 
software packages developed by P. Montegriffo at the Osservatorio 
Astronomico di Bologna. We thank Ivo Saviane for kindly providing the 
photometric catalog of 47~Tuc in electronic form. E.~P. aknowledges the
ESO Studenship Programme.


 
\begin{thebibliography}{} 
 
 
\bibitem[Alard(2001)]{ala01} Alard, C.\ 2001, \aap, 377, 389 
\bibitem[Bellazzini et al.(1999)]{b99b}Bellazzini, M., Ferraro, F.~R.; 
Buonanno, R. 1999, \mnras, 307, 619 (B99) 
\bibitem[Bellazzini et al.(2001)]{b01} Bellazzini, M., Ferraro, F.~R., 
Pancino, E.\ 2001, \mnras, 327, L15 
\bibitem[Bellazzini et al.(2002)]{b02} Bellazzini, M., Ferraro, F.~R., 
Origlia, L., Cacciari, C., Pancino, E., Monaco, L., Oliva, E.\ 2002, 
\aj, submitted 
\bibitem[Bonifacio et al.(2000)]{bo00} Bonifacio, P., Hill, 
V., Molaro, P., Pasquini, L., Di Marcantonio, P., \& Santin, P.\ 2000, 
\aap, 359, 663 
\bibitem[Buonanno et al.(1995)]{buo} Buonanno, R., Corsi, C.~E., 
Pulone, L., Pecci, F.~F., Richer, H.~B., \& Fahlman, G.~C.\ 1995, 
\aj, 109, 663
\bibitem[Cacciari, Bellazzini \& Colucci(2001)]{cbc}Cacciari, C., 
Bellazzini, M. \& Colucci, S., 2001, in Extragalactic Star Clusters, 
IAU Symp. 207, E.K. Grebel, D. Geisler and D. Minniti Eds., S. 
Francisco: ASP, in press (astro-ph/0106013) 
\bibitem[Carpenter(2001)]{car01} Carpenter, J.~M.\ 2001, \aj, 121, 2851 
\bibitem[Carretta et al.(2000)]{carr00}Carretta, E., Gratton, R.~G., 
Clementini, G., \& Fusi Pecci, F.\ 2000,\apj, 533, 215 
\bibitem[Cassisi \& Salaris(1997)]{cs97} Cassisi, S. \& Salaris, M.\
1997, \mnras, 285, 593
\bibitem[Cole(2001)]{co01} Cole, A.~A.\ 2001, \apjl, 559, L17 
\bibitem[C\^ot\'e (2000)]{cote} C\^ot\'e, P., Marzke, R.O., West, M.J., \& 
Minniti, D., 2000, \apj, 533,869 
\bibitem[Cseresnjes(2001)]{cse01} Cseresnjes, P.\ 2001, \aap, 375, 909 
\bibitem[Dean, Warren \& Cousins(1978)]{dean} Dean, J.F., Warren, P.R., 
\& Cousins, A.W., 1978, \mnras, 183, 569 
\bibitem[Dolphin(2002)]{dolphin} Dolphin, A.~E.\ 2002, \mnras, 332, 91 
\bibitem[Elias et al.(1982)]{elias} Elias, J.~H., Frogel, J.~A., 
Matthews, K., \& Neugebauer, G.\ 1982, \aj, 87, 1029 
\bibitem[Ferraro et al.(1999)]{f99} Ferraro, F.~R., Messineo, M., 
Fusi Pecci, F., de Palo, M.~A., Straniero, O., Chieffi, A., Limongi, M. 
\ 1999, \aj, 118, 1738 
\bibitem[Ferraro et al.(2000)]{f00} Ferraro, F.R., Montegriffo, P., 
Origlia, L., \& Fusi Pecci, F., 2000, \aj, 119, 1282 
\bibitem[Fusi Pecci et al.(1990)]{fp90} Fusi Pecci, F., Ferraro, F.~R., 
Crocker, D.~A., Rood, R.~T., Buonanno, R.\ 1990, \aap, 238, 95 
\bibitem[Ibata et al.(1994)]{igi} Ibata, R.~A., Gilmore, G., \& Irwin, 
M.~J.\ 1994, \nat, 370, 194 
\bibitem[Ibata et al.(1997)]{iba97} Ibata, R.~A., Wyse, R.~F.~G., 
Gilmore, G., Irwin, M.~J., \& Suntzeff, N.~B.\ 1997, \aj, 113, 634 
\bibitem[Iben(1968)]{iben} Iben, I.~Jr.\ 1968, \nat, 220, 143 
\bibitem[King, Da Costa, \& Demarque(1984)]{king} King, C.~R., Da Costa, 
 G.~S., \& Demarque, P.\ 1984, \baas, 16, 529 
\bibitem[Layden \& Sarajedini(2000)]{ls00}Layden, A.C. \& Sarajedini, A. 
2000, \aj, 119, 1760 (LS00) 
\bibitem[Majewski et al.(1999)]{maj99} Majewski, S.~R., Siegel, M.~H., 
Patterson, R.~J., \& Rood, R.~T.\ 1999, \apjl, 520, L33 
\bibitem[Marconi et al.(1998)]{marc}Marconi, G., Buonanno, R., 
Castellani, M., Iannicola, G., Molaro, P., Pasquini, L., \& Pulone, 
L., 1998, \aap, 330, 453 
\bibitem[Newberg et al.(2002)]{new02} Newberg, H.~J.~et al.\ 2002, 
\apj, 569, 245 
\bibitem[Ortolani et al.(1995)]{orto} Ortolani, S., Renzini, 
A., Gilmozzi, R., Marconi, G., Barbuy, B., Bica, E., \& Rich, R.~M.\ 1995, 
\nat, 377, 701 
\bibitem[Rosenberg et al.(2000)]{ros00} Rosenberg, A., Piotto, G., 
Saviane, I., Aparicio, A.\ 2000, \aaps, 144, 5 
\bibitem[Sarajedini \& Layden(1995)]{sl95} Sarajedini, A. \& Layden, 
A.~C.\ 1995, \aj, 109, 1086 
\bibitem[Savage \& Mathis(1979)]{sama} Savage, B.~D.~\& 
Mathis, J.~S.\ 1979, \araa, 17, 73 
\bibitem[Saviane, Rosenberg, Piotto, \& Aparicio(2000)]{sav} 
Saviane, I., Rosenberg, A., Piotto, G., \& Aparicio, A.\ 2000, \aap, 355, 966 
\bibitem[Searle \& Zinn(1978)]{sz78} Searle, L. \& Zinn, R.\ 1978, \apj, 
225, 357 
\bibitem[Schechter et al.(1993)]{dophot} Shechter, P., Mateo, M., \& 
Saha, A.\ 1993, \pasp, 105, 1342 
\bibitem[Smecker-Hane \& McWilliam(2002)]{sm02}Smecker-Hane, T.~A., \& 
McWilliam, A., 2002, ApJ, submitted (astro-ph/0205411) 
\bibitem[Thomas(1967)]{thomas} Thomas, H.-C. \ 1967, Z. Astrophys, 67, 420 
\bibitem[Walker(1994)]{wa} Walker, A.~R.\ 1994, \aj, 108, 
555 
\bibitem[Zoccali et al.(1999)]{z99} Zoccali, M., Cassisi, S., Piotto, G., 
Bono, G., \& Salaris, M.\ 1999, \apjl, 518, L49 
 
 
\end{thebibliography}
\end{document}